\documentclass[twocolumn,pre,showpacs]{revtex4}
\usepackage{graphicx}
\usepackage{times}

\begin{document}

\title{Robustness of complex many-body networks: Novel perspective in 2D metal-insulator transition}

\author{Chung-Pin Chou}

\affiliation{Beijing Computational Science Research Center, Beijing
100084, China}

\begin{abstract}
We present a novel theoretical framework established by complex
network analysis for understanding the phase transition beyond the
Landau symmetry breaking paradigm. In this paper we take a
two-dimensional metal-insulator transition driven by electron
correlations for example. Passing through the transition point, we
find a hidden symmetry broken in the network space, which is
invisible in real space. This symmetry is nothing but a kind of
robustness of the network to random failures. We then show that a
network quantity, small-worldness, is capable of identifying the
phase transition with/without any symmetry breaking in the real
space and behaving as a new order parameter in the network space. We
demonstrate that whether or not the symmetry is broken in real space
a variety of phase transitions in condensed matters can be
characterized by the hidden symmetry breaking in the weighted
network, that is to say, a decline in network robustness.
\end{abstract}


\maketitle

\section{Introduction}\label{intro}
Phases of matter are conventionally distinguished by using Landau's
approach, which characterizes phases in terms of underlying
symmetries that are spontaneously broken. The information we need to
understand phase transitions is encoded in appropriate correlation
functions, e.g., the correlation length would diverge close to a
quantum critical point. Particularly, the low-lying excitations and
the long-distance behavior of the correlations near the critical
phase are believed to be well described by a quantum field theory. A
major problem is, however, that in some cases it is unclear how to
extract important information from the correlation functions if
these phases do not break any symmetry, such as a nonmagnetic Mott
transition \cite{MottBook74}.

Owing to the breakthrough in the control of ultracold atoms on
optical lattices, Mott's original ideas about metal-insulator
transitions have been realized in recent years
\cite{GreinerNat02,StoferlePRL04,SpielmanPRL07}. His brilliant idea
is well known as the Mott transition that a metal-insulator
transition is driven by electron correlations. Although Mott
insulators are usually related to antiferromagnetism, the Mott
transition can also take place in the absence of long-range magnetic
order. In the past years, much progress has been made from both
theoretical and experimental sides in understanding Mott transitions
\cite{ImadaRMP98}. A difficulty is the absence of the order
parameter to characterize the critical behavior of the observables
for the Mott transition not accompanied by the onset of
antiferromagnetic order.

There are two earlier concepts as to what should be the order
parameter to describe the physics around the Mott transition point
from the metallic side. One is based on the disappearance of the
Fermi liquid quasiparticles introduced by Brinkman and Rice
\cite{BrinkmanPRB70}. However, the Brinkman-Rice transition actually
exists only in infinite dimensions. Notice that the Mott transition
still can be characterized by the charge stiffness, which is the
response to a twist, while no symmetry is broken here
\cite{ContinentinoBook01}. The other concept is that the
metal-insulator transition can be viewed as a condensation of doubly
occupied sites (doublons), namely, the reduction of the number of
carriers in the insulating side \cite{CastellaniPRL79}. For
simplicity, we will use the latter to study the Mott transition, and
also assume the trial wave function with only nearest-neighbor
attractive correlations between a doublon and an empty site (holon)
in variational calculations.

In this paper, we propose the weighted network constructed by the
correlation between a doublon and a holon in the two-dimensional
(2D) Hubbard model on a square lattice. Complex network theory has
become one of the most powerful frameworks for understanding the
network structures of many real systems
\cite{AlbertRMP02,DorogovtsevAIP02,NewmanSIAM03,BoccalettiPR06,DorogovtsevRMP08}.
According to graph theory, the elements of the system often are
called nodes and the relationships between them, which a weight is
associated with, are called links. The weight of the network link
contains useful information about topologies of the weighted network
\cite{GradyNatC12}. Decades ago, this unnoticed idea constructing a
weighted network from condensed matters had been proposed in quantum
Hall systems \cite{SenthilPRL99}. However, we here define the link
weight carrying important information as doublon-holon binding
correlations in the Hubbard model.

\begin{figure}[b]
\center
\includegraphics[height=1.4in,width=3in]{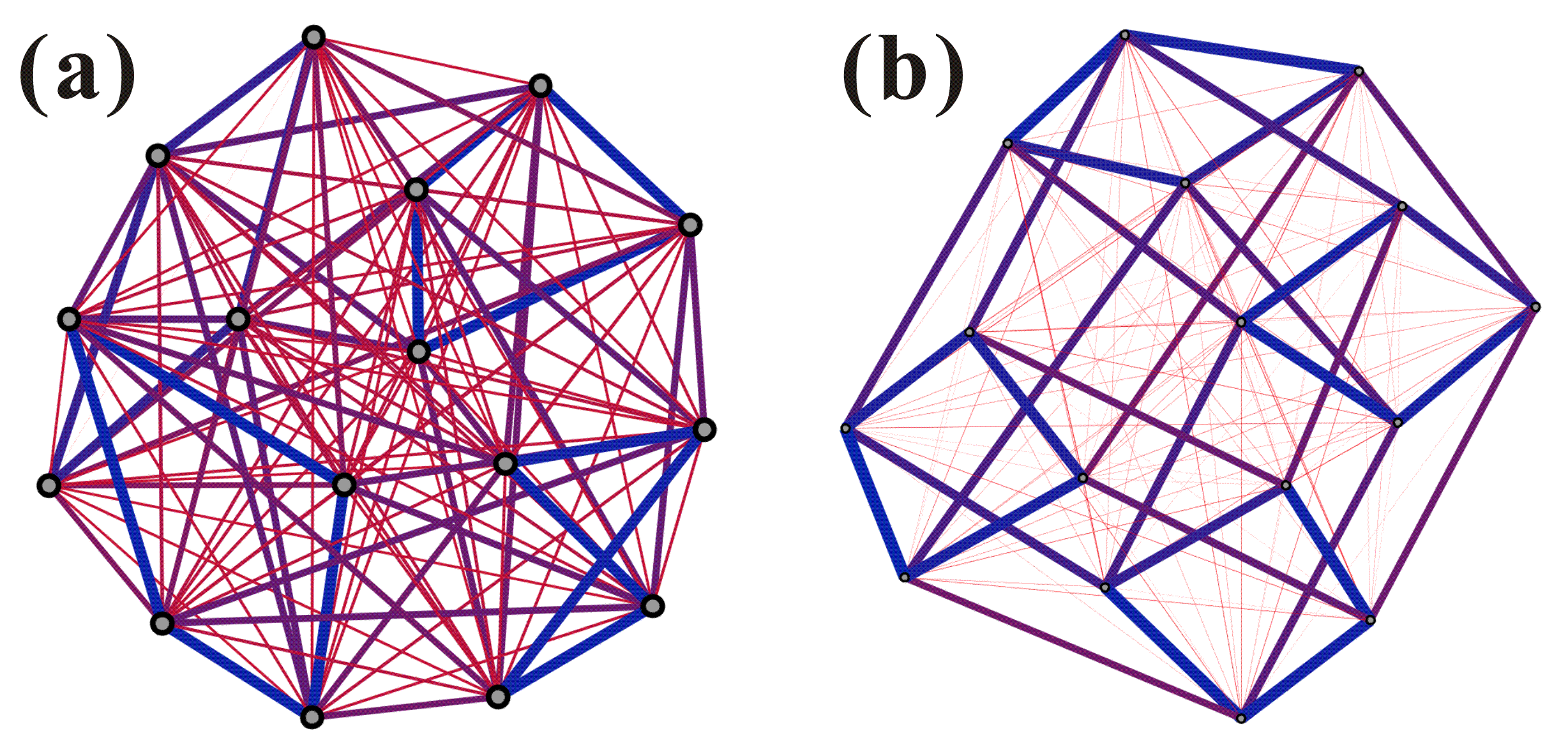}
\caption{Network representations of the 2D Hubbard model at
interaction strength (a) $U=0$ and (b) $U=12$. The lattice size
$N=4\times4$. The thickness of links represents the magnitude of the
doublon-holon correlation function. Color scale: Blue (Red)
indicates the largest (smallest) link weights.}\label{fig1}
\end{figure}

In Fig.\ref{fig1}, one can see the network topologies corresponding
to the metallic and insulating regimes (see the numerical details in
Sec.\ref{MaM}). The weighted network of the insulating state
exhibits that there are only few links with the strongest weight
called "highways" in the network, which looks very different from
the metallic state. This observation reminds us of a well-known fact
in real-world networks that heterogeneous networks are substantially
more robust to random attacks or failures than homogeneous networks
\cite{AlbertNat00,WangCM05}. The robustness of the network from its
heterogeneity seems to indicate a hidden symmetry in network space.
More precisely, the symmetry describes a phenomenon that the network
function and structure almost remain unchanged or invariant under
random removal of its links. Thus it could allow us to define an
order parameter in the Mott transition by using appropriate network
measures.

The small-world network containing both strong clustering and
shortest path length \cite{WattsNat98} has been considered to be a
well-established fact in many real-world networks
\cite{AlbertNat99,NewmanPNAS01,DoddsSci03,BackstromProc12}. Recently
the small-world idea has been successfully implemented in several
statistical models as well \cite{CPCArXiv13}. In order to quantify
the small-world network, we further propose the small-worldness as
the order parameter indicating the hidden symmetry broken in network
space, which would suddenly change while the Hubbard system
undergoes a Mott transition. We find that it is able to extract the
same Mott transition point obtained by conventional quantities, such
as charge stiffness. Based on this result as well as previous
evidences obtained from different models \cite{CPCArXiv13}, we
demonstrate that the hidden symmetry related to resilience of
networks can provide us another route to identify various many-body
phase transitions with/without Landau symmetry breakings.

\section{Model and method}\label{MaM}
We begin with the 2D half-filled Hubbard model on a square lattice,
which is relevant to correlated electron materials
\cite{GebhardBook97}:
\begin{eqnarray}
\hat{H}&=&\hat{H}_{t}+\hat{H}_{U}\\\nonumber &=&-\sum_{\langle
i,j\rangle,\sigma}\left(c^{\dag}_{i\sigma}c_{j\sigma}+H.c.\right)+U\sum_{i}\hat{d}_{i},\label{e:equ1}
\end{eqnarray}
where $\hat{d}_{i}=n_{i\uparrow}n_{i\downarrow}$ and
$n_{i\sigma}=c^{\dag}_{i\sigma}c_{i\sigma}$ with spin $\sigma$. In
the following, we express the trial ground state in the finite
system of $N$ sites with (anti-)periodic boundary condition along
the $\hat{x}$ ($\hat{y}$) direction to satisfy the closed-shell
condition.

For our numerical simulations, we use the variational Monte Carlo
(VMC) approach, which is based on the well-known Gutzwiller wave
function \cite{GutzwillerPRL63}
\begin{eqnarray}
|\Psi_{G}\rangle=\hat{P}_{G}|\Psi_{FL}\rangle,\label{e:equ2}
\end{eqnarray}
where $|\Psi_{FL}\rangle$ is the Fermi liquid and
$\hat{P}_{G}=g^{\sum_{i}\hat{d}_{i}}$. Here $g$ is the parameter to
adjust the number of doublons in the system. However, it is
impossible to realize the nonmagnetic Mott transition by just using
$|\Psi_{G}\rangle$. To demonstrate this transition, the simplest
doublon-holon binding correlation must be considered
\cite{CastellaniPRL79,YokoyamaPTP02},
\begin{eqnarray}
|\Psi_{DH}\rangle=\prod_{i}\left(1-\mu\hat{Q}_{i}\right)|\Psi_{G}\rangle,\\
\hat{Q}_{i}=\hat{d}_{i}\prod_{\delta}\left(1-\hat{h}_{i+\delta}\right)+\hat{h}_{i}\prod_{\delta}\left(1-\hat{d}_{i+\delta}\right).\label{e:equ3}
\end{eqnarray}
Here $\hat{h}_{i}=(1-n_{i\uparrow})(1-n_{i\downarrow})$ and $\delta$
runs over the nearest neighbors of the site $i$. $\mu$ is the
parameter to control the number of isolated doublons and holons. If
$\mu=0$, $|\Psi_{DH}\rangle$ will be back to $|\Psi_{G}\rangle$
which is metallic for $U<\infty$. On the other side, $\mu=1$,
$|\Psi_{DH}\rangle$ becomes a completely doublon-holon bound state
that seems to be insulating. It has been numerically proven that
$|\Psi_{DH}\rangle$ is enough to describe the behavior of the Mott
transition in VMC studies, even though we still need to consider
more correlation factors to correctly capture the transition point
\cite{MiyagawaJPSJ11}. Thus, it is reasonable to illustrate the
occurrence of a nonmagnetic Mott transition by using the
nearest-neighbor doublon-holon bound state.

In the language of network analysis, each network of $N$ nodes can
be described by the $N\times N$ adjacency matrix $\hat{A}$
\cite{AlbertRMP02}. A number of real systems, {\it e.g.} social
networks, transportation networks, biological networks and so on,
are better captured by weighted networks in which links use the
weight to quantify their strengths. Here we consider lattice sites
as nodes of the weighted network of which each weighted link between
nodes $i$ and $j$ is expressed by the element of the adjacency
matrix $\hat{A}_{ij}$. The link should carry the weight containing
information about the relationship between electrons in the Hubbard
system. According to the trial wave function we use, the weight of
the network link is naively defined by the doublon-holon binding
correlation. Namely, the lattice now breaks into $N$ single sites
called nodes, and then each pair of nodes $i$ and $j$ is reconnected
by using the doublon-holon correlation function given by
$\langle\hat{d}_{i}\hat{h}_{j}\rangle$.

Many of real networks have the property of relatively short average
path length that is a shortest route running along the links of a
network. In other words, most nodes may not be neighbors but can
reach each other by a small number of steps. This is called
small-world phenomena \cite{WattsBook99}. However, the
Erd\"{o}s-R\'{e}nyi random networks showing the small-world effect
fail to reproduce some important features of real networks, e.g.
clustering. A small-world network including not only strong
clustering but also short path length has thus been introduced to
describe real networks by Watts and Strogatz \cite{WattsNat98}.
Instead of the weighted clustering coefficient $\langle c\rangle$
\cite{OnnelaPRE05} and the average path length $\langle d\rangle$
\cite{NewmanBook10} commonly used in network analysis, an
alternative measurement of the small-world property called
"small-worldness" has been recently proposed \cite{HumphriesPLos08}.
The definition is based on the maximal tradeoff between high
clustering and short path length. We can further define
small-worldness as
\begin{eqnarray}
\langle s\rangle\equiv\frac{\langle c\rangle}{\langle
d\rangle}.\label{e:equ4}
\end{eqnarray}
One can see more details in Ref.~\cite{CPCArXiv13}. A network with
larger $\langle s\rangle$ has a higher small-world level. Since the
small-worldness simultaneously contains the information about both
locality (weighted clustering coefficient $\langle c\rangle$) and
non-locality (average path length $\langle d\rangle$) in network
space, we can expect that the small-worldness is very suitable for
describing the universal critical properties, especially the phase
transitions without local order parameters. In fact, we have shown
that the small-worldness indeed can behave like an order parameter
in either continuous or topological phase transitions in our
previous works \cite{CPCArXiv13}.

\begin{figure}[t]
\center
\includegraphics[height=2.5in,width=3in]{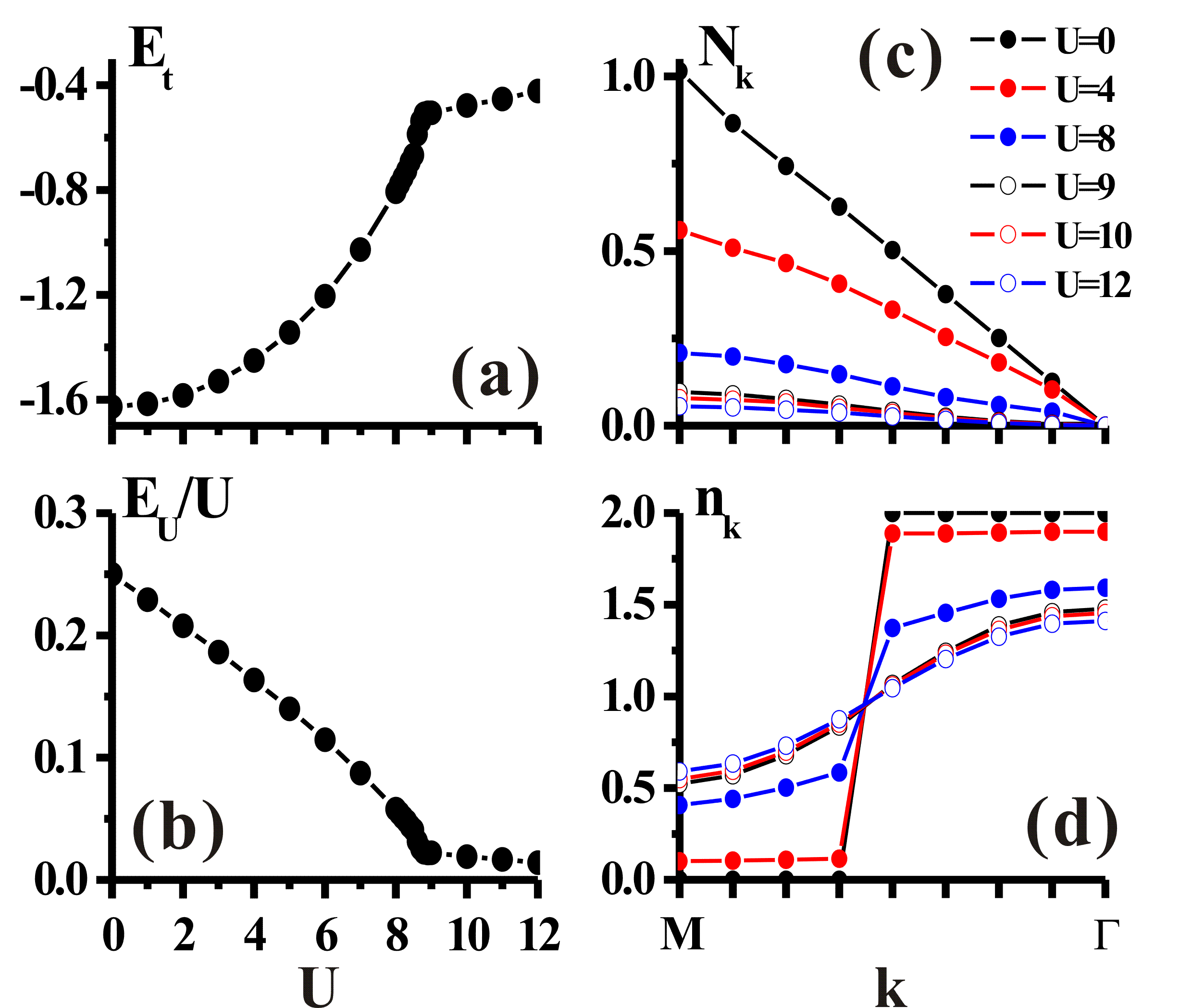}
\caption{(a) The kinetic energy $E_{t}$ and (b) the interaction
energy $E_{U}/U$ of the 2D Hubbard Hamiltonian as a function of $U$
for the lattice size $N=16\times16$. (c) The charge structure factor
$N_{k}$ and (d) the momentum distribution function $n_{k}$ along the
nodal direction from $M$ $(\pi,\pi)$ to $\Gamma$ $(0,0)$ for
different $U$.}\label{fig2}
\end{figure}

\section{Mott transition of physical observables}\label{1dq}
We now recall some facts about the nonmagnetic Mott transition
arising from the doublon-holon binding effect \cite{MiyagawaJPSJ11}.
The short-range doublon-holon binding correlation to describe the
essence of the first-order Mott transition has been confirmed by
several VMC calculations \cite{YokoyamaJPSJ90,YokoyamaJPSJ04}. A
recent VMC study also concludes that the size of the doublon-holon
bound state in the insulating phase is not beyond the next nearest
neighbors \cite{MiyagawaJPSJ11}. This result thus motivates us to
simply consider the nearest-neighbor doublon-holon binding
correlation in the trial wave function $|\Psi_{DH}\rangle$. However,
the critical issue is that the proposals on the order parameter are
still controversial even if the metal-insulator transition is well
defined. In the following, we therefore examine several conventional
physical observables to detect the critical point.

Figure \ref{fig2}(a) and (b) show the kinetic energy
$E_{t}$($\equiv\langle\hat{H}_{t}\rangle$) and the interaction
energy $E_{U}/U$($\equiv\langle\hat{H}_{U}\rangle/U$) which is
substantially the doublon density. It is obvious that both the
kinetic energy and the doublon density have a discontinuity at the
critical point $U_{c}$ that has been indicated by the other VMC
approach ($U_{c}=8.575$) \cite{MiyagawaJPSJ11}. This is a typical
sign of a first-order transition. Nevertheless, we cannot do
anything but further compute the maximum decreasing (minimum
increasing) rate of the interaction energy $-\frac{\partial
E_{U}}{\partial U}$ (the kinetic energy $\frac{\partial
E_{t}}{\partial U}$) if we want to precisely know the value of
$U_{c}$.

To estimate the critical point more accurately, we may need to
consider other physical quantities. First, we start the charge
density correlation function in momentum space called charge
structure factor,
\begin{eqnarray}
N_{k}=\frac{1}{N}\sum_{i,j}\langle n_{i}n_{j}\rangle
e^{ik\cdot(R_{i}-R_{j})}-n^{2},\label{e:equ5}
\end{eqnarray}
where $n_{i}=\sum_{\sigma}n_{i\sigma}$ and $n$ electron density. One
can see that the important information about the metal-insulator
transition is concealed in the charge structure factor $N_{k}$.
Within the variational theory, $N_{k}\propto|k|$ for
$|k|\rightarrow0$ if there is no charge gap, whereas
$N_{k}\propto|k|^{2}$ if a charge gap opens. As illustrated in
Fig.\ref{fig2}(c), the momentum dependence of $N_{k}$ near the
$\Gamma$ point abruptly changes between $U=8$ and $9$, which
coincides with the critical region determined above. Even so, the
way to distinguish the insulating phase from the metallic phase will
need much more efforts on the interpolation between lattice points.

Then, we consider the momentum distribution function,
$n_{k}=\sum_{\sigma}\langle c^{\dag}_{k\sigma}c_{k\sigma}\rangle$.
According to Fermi liquid theory, the jump of $n_{k}$ near the Fermi
surface, namely, the quasiparticle renormalization factor, can be
used to identify the metal-insulator transition. It is also in
connection with the charge stiffness. We show a clear discontinuity
of $n_{k}$ around the Fermi surface for small $U$ in
Fig.\ref{fig2}(d), thus suggesting that it is the metallic phase. As
further increasing $U$, the jump is evidently reduced and remains
very small for large $U$($>9$). The small residual value of the jump
is basically due to the finite size effect. That means we need to
make more endeavors to indicate the critical point $U_{c}$ estimated
from the extrapolation to where the jump vanishes. So far, these
observables we discuss above still cannot be thought of as the order
parameter in the Mott transition even though they are able to
distinguish the metallic and insulating phases.

\begin{figure}[t]
\center
\includegraphics[height=2in,width=3in]{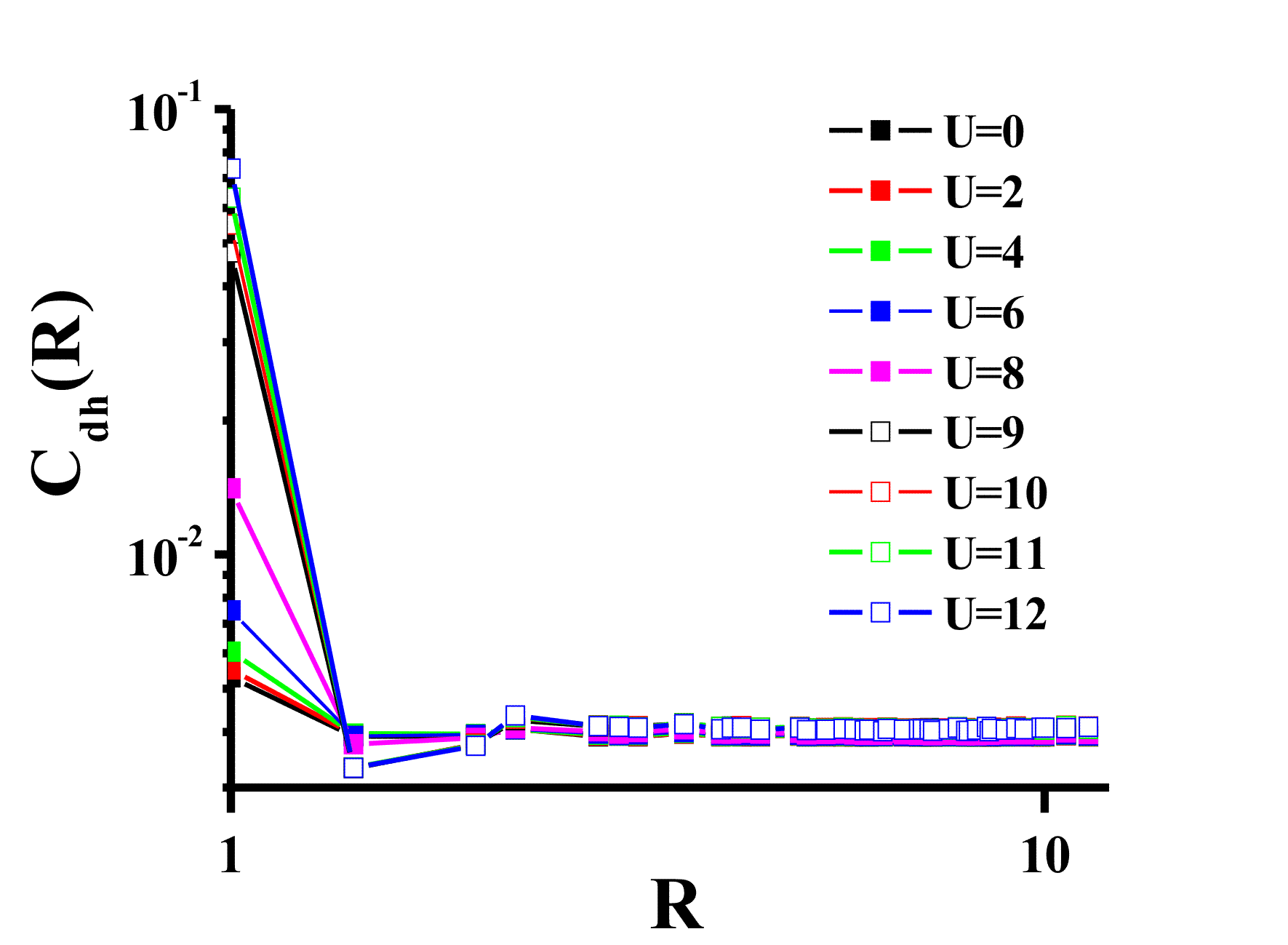}
\caption{The doublon-holon correlation function $C_{dh}(R)$ of the
2D Hubbard model of lattice size $N=16\times16$ vs distances $R$ for
various $U$.}\label{fig3}
\end{figure}

\section{Network analysis}\label{2dc}
In general, if a system shows the conventional long-range order,
like ferromagnetism or superconductivity, the corresponding
correlation function will be nonzero at its tail that can be used to
define the order parameter. Otherwise, the correlation function
would decay to zero at long distance for a system does not break any
symmetry. Here we would conjecture that it is impossible to find the
order parameter derived from the doublon-holon correlation function
because of unbroken symmetry in the nonmagnetic Mott transition. In
what follows we shall illustrate a symmetry breaking unseen in real
space.

From the doublon-holon binding effect leading to the nonmagnetic
Mott transition we can reasonably infer that the correlation between
a doublon and a holon provides a sufficient amount of data about the
critical point $U_{c}$. We define the doublon-holon correlation
function as follows,
\begin{eqnarray}
C_{dh}(R)=\frac{1}{N}\sum_{i}\langle\hat{d}_{i}\hat{h}_{i+R}\rangle.\label{e:equ6}
\end{eqnarray}
In Fig.\ref{fig3}, we show how the doublon-holon correlation
function changes as the Mott transition occurs. For $U=0$, the
Fermi-liquid ground state would not present unusual relationship
between a doublon and a holon. As increasing $U$, a doublon starts
to attract one holon at the nearest-neighbor site so that the
doublon-holon correlation function slowly enhances only at $R=1$.
When $U>U_{c}$, $C_{dh}(R)$ rapidly increases at $R=1$ as well as
$C_{dh}(R=1)\gg C_{dh}(R\neq1)$ since the doublon-holon bound state
begins to develop in the insulating phase. A key observation from
Fig.\ref{fig3} is that while the system enters the insulating
regime, the short-range parts ($R<2$) of the doublon-holon
correlation function immediately decay with distances but the
long-range parts still remain similar to the metallic phase. In
order to search a possible order parameter in network space, we will
take advantage of this property of which the correlation function
behaves differently for small and large $U$.

We now investigate how the network techniques perform in the face of
the metal-insulator transition. Based on the observation on the
doublon-holon correlation, a straightforward definition for the
elements of the adjacency matrix $\hat{A}_{ij}$ is the magnitude of
the normalized doublon-holon correlation function between the
lattice sites $i$ and $j$,
\begin{eqnarray}
\hat{A}_{ij}=\frac{|\langle\hat{d}_{i}\hat{h}_{j}\rangle|}{max(|\langle\hat{d}_{i}\hat{h}_{j}\rangle|)}.\label{e:equ7}
\end{eqnarray}
The complex topology of the weighted network extracted from
$\hat{A}_{ij}$ has been shown in Fig.\ref{fig1} for the small
lattice. Obviously, the weight of network links exhibits different
distributions for the metallic and insulating phases even though all
nodes in the network are completely connected. It would be just a
trivial complete network if the network were binary. Next, we need
to collect useful information from a large volume of network data.

\begin{figure}[t]
\center
\includegraphics[height=1.2in,width=3.4in]{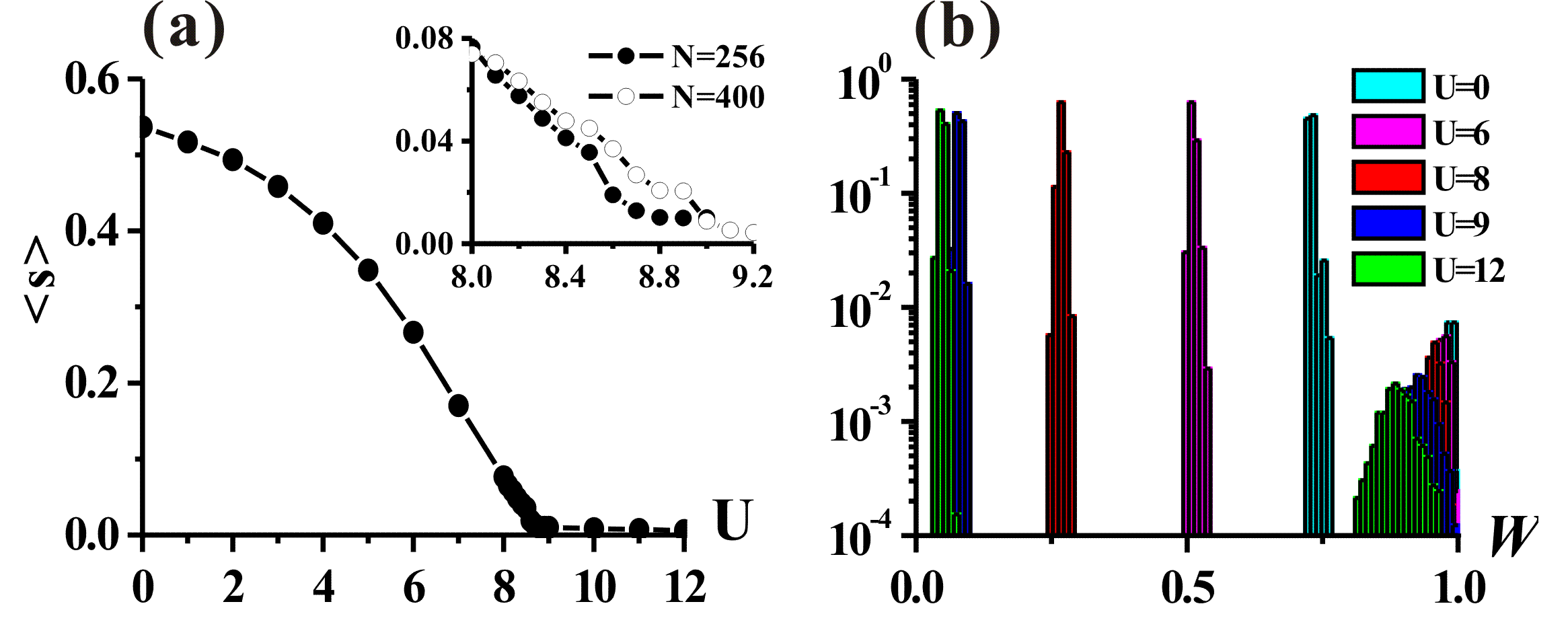}
\caption{(a) Small-worldness $\langle s\rangle$ of the 2D Hubbard
model as a function of $U$ for $N=16\times16$. The inset shows the
magnification near the critical point and includes the data points
in a larger lattice system ($N=20\times20$). (b) The probability
distribution of the link weight $W$ for different $U$. The bin size
is chosen as $0.01$ for clear demonstrations. The lattice size
$N=16\times16$.}\label{fig4}
\end{figure}

In our previous studies \cite{CPCArXiv13}, we have confirmed that
the small-worldness $\langle s\rangle$ defined in Eq.(\ref{e:equ4})
has the ability to detect the second-order and topological phase
transitions. Here we further examine the feasibility of the
small-worldness in the first-order phase transition without any
symmetry breaking. In Fig.\ref{fig4}(a), we illustrate the critical
behavior of the small-worldness in the 2D Hubbard model. One can see
that the $U-$dependence of the small-worldness behaves like an order
parameter and drops down to almost zero near the critical point
$U_{c}$ in the finite system of size $N=16\times16$. In the inset of
Fig.\ref{fig4}(a), we find that the critical point is indicated by
the sudden reduction of $\langle s\rangle$ at $U_{c}$, in agreement
with those results obtained above and in Ref.~\cite{MiyagawaJPSJ11}.
Notice that in a larger lattice of size $N=20\times20$ the
small-worldness for $U>U_{c}$($\sim8.9$) is further reduced to zero.
However, it is rather difficult to estimate $U_{c}$ by using
finite-size calculations since the Mott critical value of the
doublon-holon binding state would become larger as increasing
lattice size. In any case, this agreement convinces us that the
small-worldness can be used as another quantity to characterize the
nonmagnetic Mott transition instead of the charge stiffness.

To understand this result, we need to show how the weight
distribution of network links enables the complex weighted network
to illustrate the Mott transition. Fig.\ref{fig4}(b) shows that the
weights of network links display two bounded distributions with the
asymmetric bimodal shape. The bimodal shape is an artifact, directly
associated with the broken $C_{4}$ rotational symmetry by
anti-periodic boundary condition along the $\hat{y}$ direction. At
$U=0$, few links have the strongest weight ($W\sim1$) from which the
weights for most links are not far away ($W\sim0.75$). Thus, if we
were to ignore the bimodal shape for the time being, the weighted
network for $U=0$ would exhibit the homogeneous distribution around
$W=1$, corresponding to the strongest small-world network whose most
links resemble the highway in road networks.

\begin{table*}
\caption{Examples of symmetry breakings and their corresponding
order parameters.}
\begin{ruledtabular}
\begin{tabular}{llll}
Normal phase&Condensed phase&Broken symmetry&Order parameter\\
\hline
Paramagnet&Ferromagnet&Spin rotation&Magnetization\\
Metal&Superconductor&Global gauge&Pair condensate\\
Heterogeneous network &Homogeneous network&Network robustness&Small-worldness\\
\end{tabular}\label{tabl1}
\end{ruledtabular}
\end{table*}

As further increasing $U$, the weights for most links begin to move
away from the largest weight ($W=1$) and hence the distributions
become more heterogeneous, leading to the reduction of the
small-worldness as shown in Fig.\ref{fig4}(a). When $U>U_{c}$, most
links have very small weights ($W\sim0$) and few links with the
stronger weights ($W\sim0.9$) show the broad distribution. The
phenomena that most links belong to slow traffic lanes would result
in almost zero small-worldness (weak clustering and long path). The
same reasoning from the weight distribution of network links could
be also applied to other many-body systems without local order
parameters in real space. Hence this result strongly suggests that
the small-worldness can be considered as a new order parameter in
the network representation to capture the nonmagnetic Mott
transition.

We have to mention a point now in passing. In the network
representation, there exists a kind of hidden symmetry corresponding
to the heterogeneous network with a broad weight distribution. This
symmetry is related to the robustness of the network that is
resilient to random attacks or failures \cite{WangCM05}. The
heterogeneity of network links implies that the weighted network is
much more robust against attacks than the homogeneous networks,
giving rise to the higher symmetry in network space. Conversely, the
homogeneity of network links means that the weighted network becomes
fragile to random attacks, and thus breaks the hidden symmetry. The
hidden symmetry breaking is in analogy with typical examples of
spontaneous symmetry breakings in condensed matter physics, such as
the ferromagnet or superconductor (see the comparison in Table
\ref{tabl1}). Therefore, we demonstrate that the small-worldness
carrying the messages about both locality (clustering) and
non-locality (path length) is able to represent the order parameter
describing the hidden symmetry broken in network space.

\section{Conclusion}\label{conc}
In this work we have addressed how to read useful information from
the doublon-holon correlation function to identify the nonmagnetic
Mott critical point by using complex network analysis. We have
compared several conventional observables describing the Mott
transition, and yet none of them can be formally treated as the
order parameter. In the 2D Hubbard model, we have illustrated that
one of many network measures, small-worldness, plays a significant
role as a general order parameter concealed in network space. The
critical point $U_{c}$ extracted from the small-worldness is very
close to the one obtained from other physical observables. Thus, it
is anticipated that the small-worldness can be applied to the
many-body systems beyond Landau symmetry-breaking theory.

The broadening of the weight distribution of network links across
the critical point is responsible for the change of the
small-worldness, which is analogous to the change of the speed limit
on a road network from the highway to the slow lane. The phenomenon
that the structure of the weighted network varies from heterogeneity
to homogeneity implies a hidden symmetry broken$-$or, to put it
another way, the reduction of the robustness to random attacks in
the network space. We have confirmed that the broken symmetry can be
successfully described by the small-worldness. The most important
finding in this work is the hidden symmetry breaking in network
space, namely, the reduction of network resilience, which has been
listed in Table \ref{tabl1}.

The weight distribution of network links is able to uncover a wealth
of complex topological information underneath correlation functions,
and further comprehends the mechanism of the phase transition
with/without any order parameter in real space. We propose that the
small-worldness determined from various weighted networks could be a
valuable tool to investigate quantum phase transitions. The finding
has been corroborated by our previous results in the 1D quantum
Ising and 2D classical XY models \cite{CPCArXiv13}. As a result,
plus the evidence given in the 2D Hubbard model, one may conjecture
that generally the point of view coming from complex network
topology can be a vital component in studying the phase transitions
in condensed matter physics.

\section*{Acknowledgments}
We thank T.-K. Lee, F. Yang, and M.-C. Chang for helpful
discussions. This work is supported by Chinese Academy of
Engineering Physics and Ministry of Science and Technology. Most of
calculations are performed in the National Center for
High-performance Computing in Taiwan.

\end{document}